\documentclass[a4paper]{cas-dc}

\usepackage[square, comma, sort&compress, numbers]{natbib}
\usepackage{subfigure}
\usepackage{booktabs}  
\usepackage{multirow}
\usepackage{algorithmic}
\usepackage{algorithm}
\usepackage{pifont}
\def\tsc#1{\csdef{#1}{\textsc{\lowercase{#1}}\xspace}}
\tsc{WGM}
\tsc{QE}

\begin{document}
\let\WriteBookmarks\relax
\def\floatpagepagefraction{1}
\def\textpagefraction{.001}

\shorttitle{Plug-in Hybrid Electric Vehicle Energy Management with Clutch Engagement Control via Continuous-Discrete Reinforcement Learning}    

\shortauthors{\textit{Energy Technol.}}  

\title [mode = title]{Plug-in Hybrid Electric Vehicle Energy Management with Clutch Engagement Control via Continuous-Discrete Reinforcement Learning}  

%
\author[]{Changfu Gong}[type=editor,
       style=chinese,
       auid=000,
       bioid=1,
       orcid=0000-0002-8805-6468]




\ead{cs_changfu@mail.scut.edu.cn}



\affiliation[]{organization={Shien-Ming Wu School of Intelligent Engineering},
            addressline={South China University of Technology}, 
            city={Guangzhou},
            postcode={511442}, 
            country={China}}

\author[]{Jinming Xu}

\ead{wi_jinming@mail.scut.edu.cn}



\author[]{Yuan Lin}
\fnmark[*]

\ead{yuanlin@scut.edu.cn}

\cortext[1]{Corresponding author. Shien-Ming Wu School of Intelligent Engineering, South China University of Technology, Guangzhou, 511442, China}



\begin{sloppypar}

\begin{abstract}
Energy management strategy (EMS) is a key technology for plug-in hybrid electric vehicles (PHEVs). The energy management of certain series-parallel PHEVs involves the control of continuous variables, such as engine torque, and discrete variables, such as clutch engagement/disengagement. We establish a control-oriented model for a series-parallel plug-in hybrid system with clutch engagement control from the perspective of mixed-integer programming. Subsequently, we design an EMS based on continuous-discrete reinforcement learning (CDRL), which enables simultaneous output of continuous and discrete variables. During training, we introduce state-of-charge (SOC) randomization to ensure that the hybrid system exhibits optimal energy-saving performance in both high and low SOC. Finally, the effectiveness of the proposed CDRL strategy is verified by comparing EMS based on charge-depleting charge-sustaining (CD-CS) with rule-based clutch engagement control, and Dynamic Programming (DP). The simulation results show that, under a high SOC, the CDRL strategy proposed in this paper can improve energy efficiency by 8.3\% compared to CD-CS, and the energy consumption is just 6.6\% higher than the global optimum based on DP, while under a low SOC, the numbers are 4.1\% and 3.9\%, respectively.
\end{abstract}


\maketitle

\section{Introduction}\label{}
Energy conservation and reducing consumption are effective ways to achieve low-carbon development of automotive technology. Plug-in hybrid electric vehicles (PHEVs) combine the advantages of electric vehicles and traditional gasoline vehicles, which can save energy and reduce emissions while avoiding the range anxiety associated with electric vehicles \cite{zhang2022energy}. The study of EMS in PHEVs involves the coordination between electric energy and fuel, which is a crucial technology impacting the vehicle's fuel economy and emissions \cite{torres2014energy}. Therefore, a reasonable and effective EMS is crucial for improving the overall performance of PHEVs and can also contribute to achieving the sustainable development goals of the automotive industry.

\subsection{Literature review}\label{}
In order to improve the fuel economy of PHEVs, a significant amount of research has been conducted over the past few decades on EMS for PHEVs. EMS can be divided into rule-based \cite{peng2017rule}, optimization-based \cite{ye2023application}, and learning-based methods \cite{li2021adaptive, zhang2021double}. Rule-based strategies select the operating mode based on pre-defined rules and can be further divided into deterministic rule-based \cite{lin2003power} and fuzzy logic-based strategies \cite{navale2014fuzzy}. Rule-based EMS is widely used due to its simplicity and practicality, but cannot obtain the globally optimal solution \cite{liu2021driving}.

In optimization-based strategies, the PHEV's EMS is typically abstracted as a constrained nonlinear optimization problem. Optimization-based strategies can be divided into two categories: global optimization \cite{yuan2013comparative} and instantaneous optimization \cite{xie2019predictive}. Global optimization primarily encompasses DP \cite{brahma2000optimal} and Pontryagin's minimum principle (PMP) \cite{zheng2014numerical}. Based on the Bellman optimality principle, the DP algorithm uses state transition equations to solve the optimal energy allocation between the engine and battery\cite{saiteja2022critical}. The EMS based on the PMP algorithm aims to obtain the optimal control strategy by minimizing the Hamilton equation in real-time. DP and PMP methods require prior knowledge of the driving cycle, which is difficult to achieve in actual driving, so these methods are mainly used for offline optimization.

Model predictive control (MPC) \cite{guo2016optimal} and the minimum equivalent fuel consumption strategy (ECMS) \cite{yang2018driving} are typical instantaneous optimization algorithms. MPC is based on rolling optimization, which confines the optimization process to a finite predictive range. The core idea of ECMS is to use equivalent coefficients to convert the motor's power consumption into fuel consumption and solve the optimal energy allocation problem \cite{rezaei2017estimation}. Compared to global optimization, instantaneous optimization requires less storage space and computational time, and offers potential for real-time control.

Reinforcement learning (RL) has been applied to energy management in PHEVs, and the results have been promising \cite{hu2022deep}. In \cite{qi2015novel}, a Q-learning-based EMS for PHEVs was proposed, which makes decisions using a look-up Q-table. It does not need to rely on prior knowledge of future driving conditions to make optimal decisions. Simulation results show that the fuel economy of the proposed EMS is improved by 11.93\% compared with the binary mode control strategy. In \cite{kouche2018power}, a hybrid electric vehicle (HEV) EMS based on the state–action–reward–state–action (SARSA) algorithm was studied. The algorithm treats the HEV controller as a learning agent that, through trial and error, learns an optimal power management strategy between the fuel cell and the battery. Simulation results show that, compared with the Q-learning algorithm, the SARSA algorithm can reduce battery charging and discharging more effectively. Q-learning and SARSA use Q-tables to represent the values of state-action pairs. However, the state space is usually multidimensional in the complex configuration of HEV energy management tasks. The storage space required by Q-table will also be large, posing challenges in terms of memory requirements. 

Some researchers have combined deep learning (DL) with RL and proposed deep reinforcement learning (DRL) \cite{lin2022co}, which uses neural networks in place of a Q-table. Compared with the Q-learning strategy using the same model, the DQL strategy performs better in terms of training difficulty and the influence of different state variables on the Q function \cite{wu2018continuous}. The DQL algorithm performs well in dealing with discrete action space problems but is unable to solve continuous action space problems \cite{wang2022parameterized}.

To solve the problem of discrete control variables, an EMS based on the Deep Deterministic Policy Gradient (DDPG) was proposed in \cite{chen2022reinforcement}. Based on DQN, DDPG introduces a policy network to output continuous actions, avoiding discretization of the action space \cite{lin2020comparison}. Simulation results show that compared with the DQN algorithm, the DDPG algorithm converges faster and is more robust. In \cite{liu2023energy}, an EMS based on TD3 is employed, which directly outputs continuous actions through the actor network. Moreover, by introducing two sets of identical Critic networks, it addresses the issue of overestimation of the Q-function that may lead to training instability in the DDPG algorithm. Simulation results indicate that, the EMS based on the TD3 algorithm exhibits better fuel economy. In conclusion, while discrete reinforcement learning and continuous reinforcement learning have their strengths, when a system involves both continuous and discrete decision variables, a single approach often struggles to fully exploit its potential.

For some hybrid systems, such as Honda's Intelligent Multi-Mode Drive (i-MMD) and BYD's Dual Mode Intelligent (DM-i), controlling clutch engagement/disengagement is essential. Optimal control of EMS in these systems requires simultaneous output of continuous variables (engine torque) and discrete variables (clutch engagement/ disengagement), which is a mixed-integer programming problem. In such hybrid systems, the engagement/disengagement of the clutch is independently controlled primarily based on vehicle speed and acceleration requirements through pre-established rules \cite{li2021multi}. While deterministic rules offer real-time responsiveness, they may sacrifice system optimality. The global optimal solution can be obtained using classical algorithms like branch and bound in mixed-integer programming \cite{richards2005mixed}. Nevertheless, when the system is highly complex and nonlinear, solving it becomes challenging and computationally intensive. CDRL directly outputs continuous and discrete actions, avoiding the computational complexity of mixed-integer programming methods, and can provide near-optimal energy management. To our best knowledge, there is currently no literature on using CDRL to optimize energy management and clutch engagement in series-parallel PHEVs simultaneously.

Furthermore, the uncertainty of the initial SOC undoubtedly increases the complexity of calculations for traditional optimization methods and poses a challenge to the generalization of DRL. However, existing literature on using DRL for energy management only trains for a single SOC value and does not consider the uncertainty of the initial SOC value \cite{du2020deep, qi2022hierarchical, xu2020ensemble, lee2020comparative, wu2019deep}. 

\subsection{Motivation and innovation}

BYD DM-i PHEVs are very popular in China, but there is a lack of literature specifically focused on the DM-i hybrid system. There is little research using CDRL for simultaneous energy management and clutch engagement/disengagement control. This article takes the DM-i hybrid system as the research object, establishes a vehicle powertrain model, and designs a CDRL algorithm that outputs both continuous and discrete variables, introducing randomization of the SOC during the training process, thereby achieving approximate optimal control in the EMS.

The main contributions of this paper are summarized as follows: 

(1) Modeling a hybrid system from the perspective of mixed-integer programming, treating clutch engagement/ disengagement as a discrete control variable to achieve simultaneous optimization of EMS and clutch engagement. 

(2) Designing a reinforcement learning algorithm Parametrized Deep Q-Network with Twin Delayed DDPG (PDQN-TD3) to train the optimal EMS of the DM-i hybrid system, achieving simultaneous selection of continuous and discrete actions. 

(3) Randomization of SOC is introduced in the training process to train an EMS that is optimal in both high and low SOC, i.e., in both charge-depleting and charge-sustaining modes.

\subsection{Organization}
The rest of this paper is organized as follows: Section 2 models the BYD DM-i PHEV. Section 3 introduces the CDRL algorithm PDQN-TD3. Section 4 describes the EMS based on the PDQN-TD3 algorithm. Section 5 shows the simulation results of PDQN-TD3 and compares it with EMS based on CD-CS and DP methods. Section 6 concludes the paper.

\section{DM-i Hybrid System Modeling}
\begin{figure}[!htb]
\centering
\includegraphics[width= 3 in]{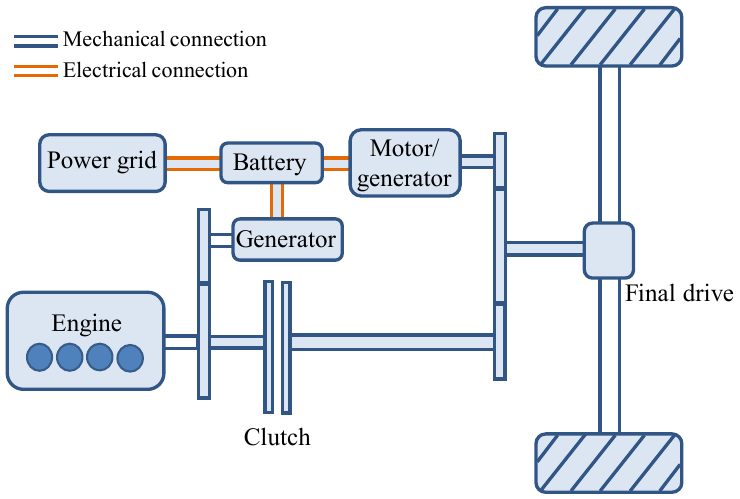}
\caption{ Configuration of the BYD DM-i PHEV.}
\label{f:DM-i}
\end{figure}

Modeling hybrid systems is the foundation for designing EMSs. In this section, we consider the DM-i PHEV system as shown in Fig. \ref{f:DM-i}, which consists of components such as an engine, a clutch, a generator, a drive motor, and a power battery. The critical component parameters are shown in Table \ref{PHEV parameters}. In this type of hybrid system, both the engine and battery serve as energy sources for the powertrain. The engine and drive motor serve as the power sources for the vehicle. By controlling the clutch engagement/disengagement, engine, and motor state, it is possible to realize five operation modes: EV mode, series mode, parallel mode, engine driving mode, and energy recovery mode.

\begin{table}[!htbp]
\renewcommand\arraystretch{1.5} 
\footnotesize
\centering
\caption{Main parameters of the PHEV.} 
\label{PHEV parameters} 
\begin{tabular}{cccc}
   \toprule
   Component & Parameter & Unit & Value \\
   \midrule
     & Mass & kg & 1500 \\
     & Windward area & m${^2}$ & 2.36 \\
     & Air drag coefficient & - & 0.28 \\
   Vehicle & Tyre radius & m & 0.3382 \\
     & Rolling resistance coefficient & - & 0.012 \\
     & EV mode gear ratio & - & 10.126 \\
     & Parallel mode gear ratio & - & 2.8 \\
     &Series mode gear ratio & - & 2.07 \\
    Engine & Maximum angular velocity & rpm & 6000 \\
     & Maximum torque & Nm & 120 \\
    Motor & Maximum angular velocity & rpm & 16000 \\
     & Maximum torque & Nm & 325 \\
    Generator & Maximum angular velocity & rpm & 13000 \\
     & Maximum torque & Nm & 110 \\
    Battery & Voltage & V & 320 \\
     & Capacity & Ah & 26 \\
   \bottomrule
\end{tabular}
\end{table}

\subsection{Vehicle dynamics}
For hybrid systems, the powertrain must obey the torque balance equation:
\begin{equation}
\label{eq:torque balance}
    T_d = T_e i_e k_c \eta_t + T_m i_m \eta_t + T_b
\end{equation}
where $T_d$ denotes the required driving torque of PHEV, $T_e$ is engine torque, $i_e$ is engine gear ratio, $k_c$ denotes clutch engagement/disengagement (with a value of either 1 or 0), $T_m$ is motor torque, $i_m$ is motor gear ratio, $T_b$ is brake torque, $\eta_t$ is the mechanical efficiencies.

According to the longitudinal dynamics equation of the vehicle, the required torque of the powertrain system is established as:
\begin{align}
\label{eq:veh_dyn}
T_d & = [ma + 0.5 C_d \rho A v^2 + \mu mg cos(\theta) + mg sin(\theta)] r\\
w_d & = \frac{v}{r}
\end{align}
where $m$ is curb weight, $a$ represents vehicle acceleration, $C_d$ is air drag coefficient, $\rho$ is air density, $A$ is the windward area, $v$ is the longitudinal vehicle velocity without regard to wind speed, $\mu$ is the rolling resistance coefficient, $g$ is the gravity acceleration, $\theta$ is the road slope, $r$ is the wheel radius, $\omega_d$ is the wheel speed.

\subsection{Engine model}
The fuel economy of engine is a key factor in evaluating the EMS of the hybrid system. In this article, an experimental modeling method is used for engine modeling. Given the engine speed and torque, the instantaneous fuel consumption rate can be obtained by interpolating the engine fuel consumption map, as shown in Fig.\ref{engine_motor_battery}(a). The engine fuel consumption per unit time is given by Eq. (\ref{fuel consumption}):
\begin{equation}
\label{fuel consumption}
\dot{m}_f = P_{e} b_e = T_e \omega_e b_e
\end{equation}
where $P_{e}$ is engine power, $b_e$ is the effective fuel consumption of the engine according to BSFC in Fig.\ref{engine_motor_battery}(a), and $\omega_e$ is engine angular velocity.

In the DM-i, the controller controls the connection and disconnection between the engine and the wheels by controlling the engagement/disengagement of the clutch. When the clutch is disengaged, the engine and wheels are decoupled, the engine speed is independent of vehicle speed, and the engine operates in the optimal working curve. When the clutch is closed, the engine and the wheel are coupled, and the engine speed adjusts according to the vehicle's speed. Therefore, the engine speed can be expressed by the following equation:
\begin{equation}
\label{w_e}
\omega_e = \omega_d i_e k_c + f(T_e)(1-k_c)
\end{equation}
where $f(T_e)$ represents the functional relationship between engine torque and speed when the engine operates along the optimal economic working curve. 

\subsection{Drive motor and generator models}

\begin{figure*}[!t]
  \centering
  \subfigure[]{\includegraphics[width=0.32\textwidth]{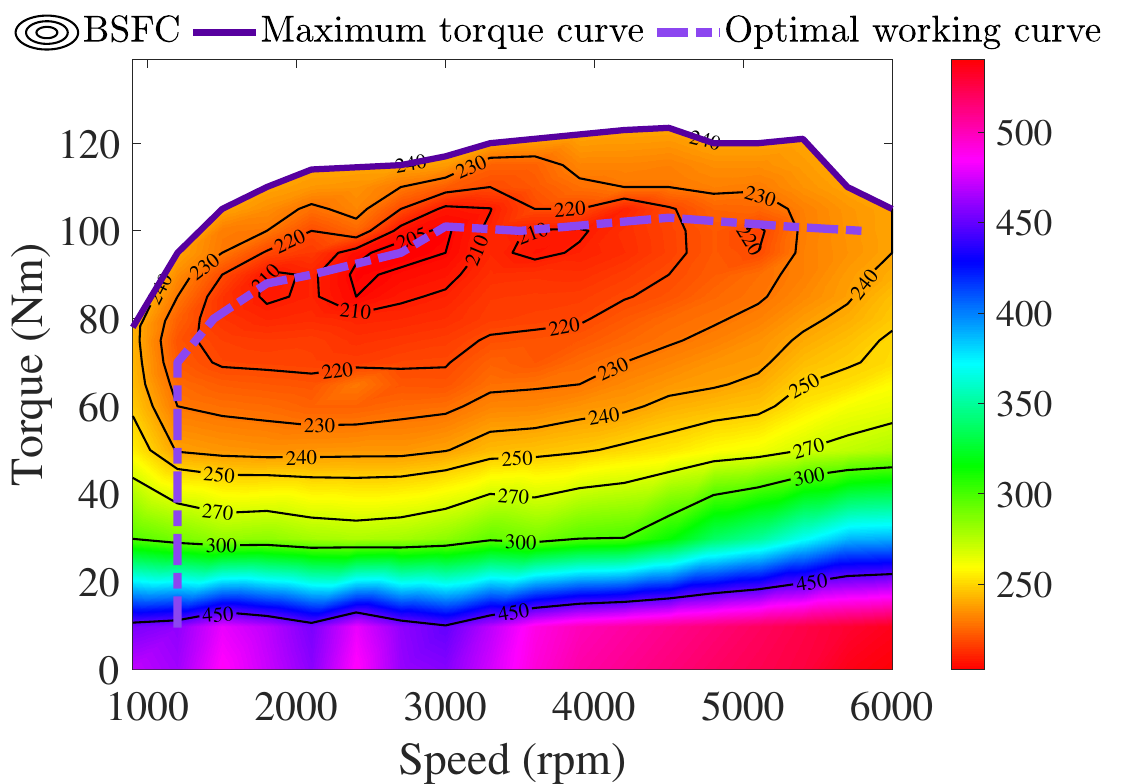}}
  \subfigure[]{\includegraphics[width=0.32\textwidth]{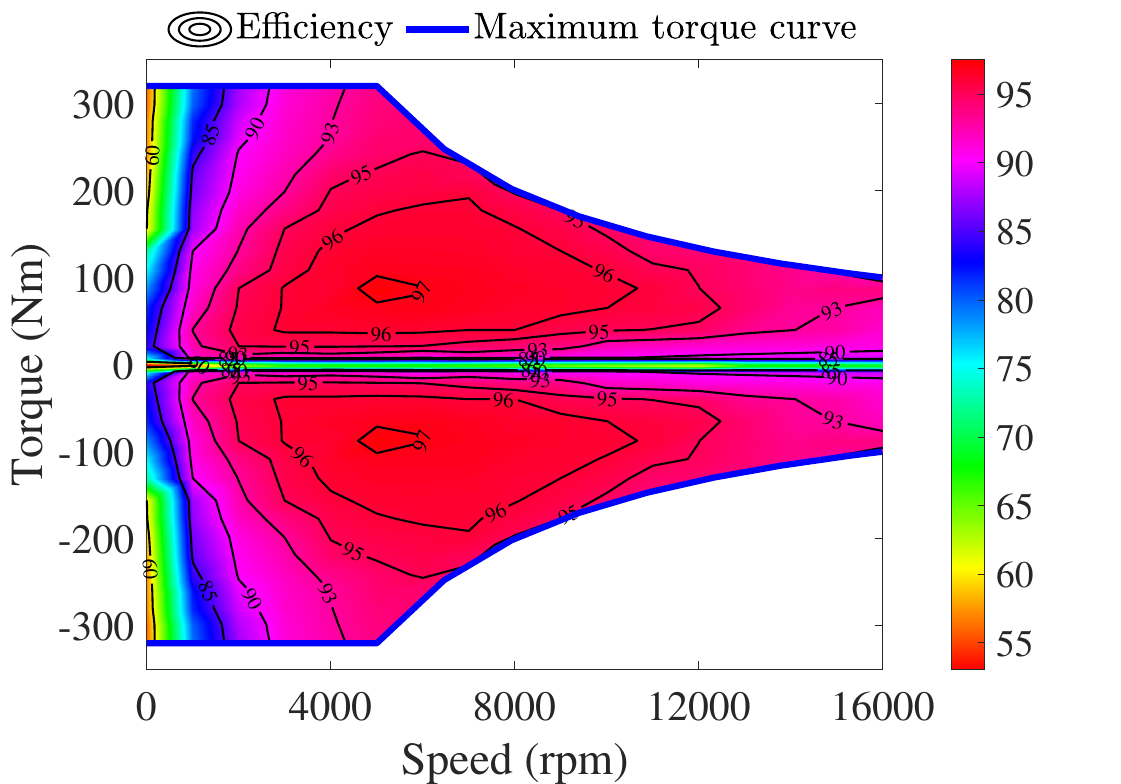}} 
  \subfigure[]{\includegraphics[width=0.32\textwidth]{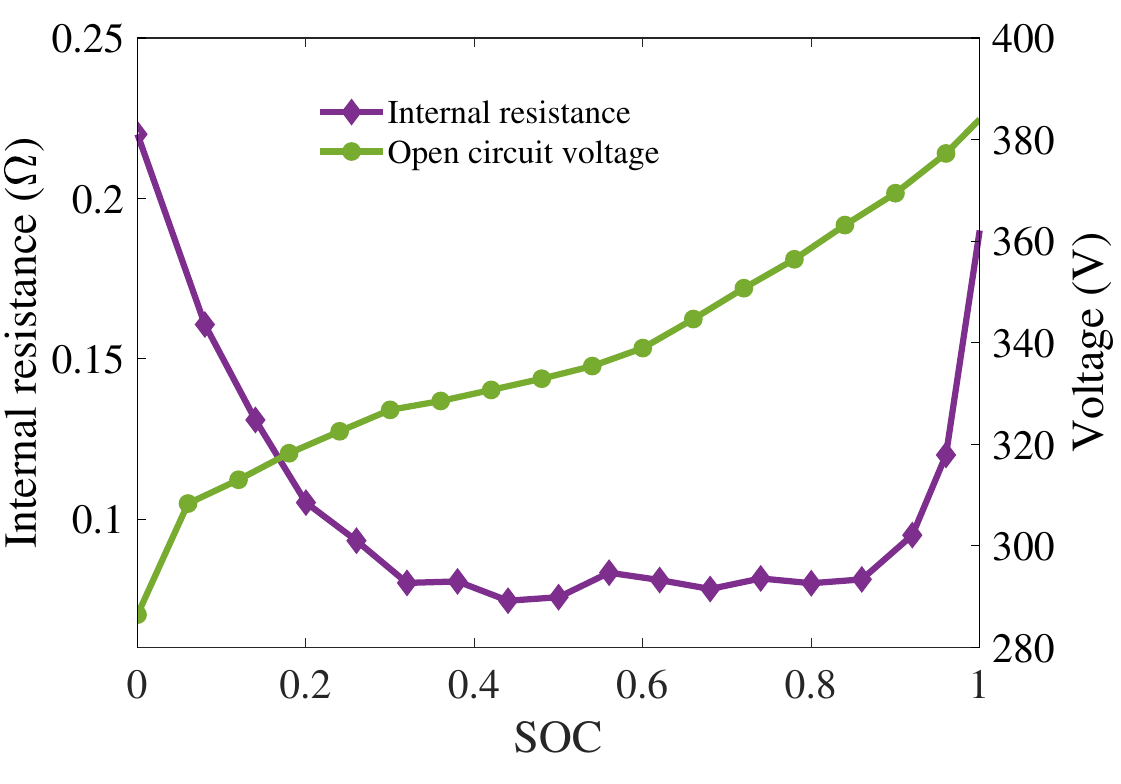}}
  \caption{(a) Engine fuel consumption map, wherein the unit of BSFC is g/kW·h. (b) Efficiency map of the driving motor. (c) Open-circuit voltage and internal resistance of battery.
  }
  \label{engine_motor_battery}
\end{figure*}

The DM-i system has two electric motors: the drive motor and the generator. The drive motor provides torque output during driving and is responsible for energy recovery during braking. The generator mainly serves as an auxiliary unit, converting mechanical energy into electrical energy, ensuring the engine's quick start, and adjusting the engine speed to maintain its operation within the economic range. The motor's map is shown in Fig. \ref{engine_motor_battery}(b). 

For the drive motor, the motor speed and motor torque are written as:
\begin{equation}
\label{motor speed}
\omega_m = \omega_d i_m 
\end{equation}
where $\omega_m$ is the driving motor speed.

To reduce the computational overhead of the RL and DP algorithm, we optimized Eq.(\ref{eq:torque balance}) by merging $T_m$ and $T_b$ into $T_{mb}$, thus eliminating the need to control $T_b$.
\begin{equation}
\label{eq:Tmb}
\begin{split}
T_{mb} = T_m i_m \eta_t + T_b = T_d  - T_e i_e k_c\eta_t
\end{split}
\end{equation}

The $T_m$ and $T_b$ can be obtained by Eq.(\ref{eq:Tmb}):
\begin{align}
\label{eq:Tmopt}
\ \ T_m =
\begin{cases}
{-f(\omega_m)}, \qquad & {\rm if} \; T_{mb} \textless -f(\omega_m) \\
{T_{mb}/i_m \eta_t}, \qquad & \rm {else}
\end{cases}
\end{align}
\begin{align}
\label{eq:Tb_opt}
\ \ T_b =
\begin{cases}
{T_{mb}-T_m i_m \eta_t = T_{mb} + f(\omega_m)i_m \eta_t},\\
 \qquad \qquad \qquad \qquad \qquad   {\rm if} \;  T_{mb} \textless -f(\omega_m) \\
{0}, \qquad \qquad \qquad \qquad  \; \; \; \; \rm {else}
\end{cases}
\end{align}
where $f(\omega_m)$ represents the maximum torque of the motor at the current speed.

For Eq. (\ref{eq:Tmopt}) and (\ref{eq:Tb_opt}), when $T_{mb}$ is less than the motor's energy recovery upper limit ($T_{mb} < -f(\omega_m)$), the vehicle enters a braking state. The motor recovers energy at maximum capacity, while the mechanical brake ($T_b$) provides the remaining braking force. When $T_{mb} > -f(\omega_m)$, the motor assumes the role of energy recovery if $T_{mb} < 0$, and driving if $T_{mb} > 0$. In both scenarios, the utilization of a mechanical brake ($T_b$) is unnecessary.

For the generator, the speed and torque are written as:
\begin{equation}
\label{generator speed}
\omega_g = \omega_e/i_g 
\end{equation}
\begin{equation}
\label{generator torque}
T_g = T_e i_g \eta_{eg} (1-k_c)
\end{equation}
where $\omega_g$ is the generator speed, $T_g$ is the generator torque, $i_g$ is the generator gear ratio, $\eta_{eg}$ is the transmission efficiency between the engine and the generator.

\subsection{Battery model}
The power battery is used to provide the electric energy required by the motor during driving and can also store the energy recovered by the motor during braking. This paper does not consider the effect of temperature on the internal characteristics of the battery but establishes the dynamic equation of SOC based on the internal resistance of the battery, as shown in the following equations:
\begin{equation}
\label{battery power}
P_b = T_m \omega_w \eta^{sgn(-T_m)}_m  + T_g \omega_g \eta_g + P_{aux}
\end{equation}
\begin{equation}
\label{battery power1}
P_b = V_{oc} I_b - R_bI^2_b
\end{equation}
\begin{equation}
\label{SOC dot}
\dot{SOC} = -\frac{V_{oc} - \sqrt{V^2_{oc} - 4 R_b P_b}}{2R_b Q_b}
\end{equation}
where $P_b$ is battery power, $\eta_m$ is the motor efficiencies, $\eta_g$ is the generator efficiency, $P_{aux}$ is the auxiliary system power consumption of the vehicle, $V_{oc}$ is the open circuit voltage, $I_b$ is battery current, $R_b$ is battery resistance. Ignoring the effects of battery aging and temperature, the relationship among $SOC$, battery internal resistance, and open circuit voltage is shown in Fig.\ref{engine_motor_battery}(c).

\section{PDQN-TD3 Continuous-Discrete Reinforcement Learning Algorithm}
Currently, RL methods predominantly focus on continuous or discrete action spaces. However, many engineering control problems involve both continuous and discrete variables, referred to as mixed action space. For example, during the driving process of PHEVs, the engine's torque is a continuous variable, while the clutch switch is a discrete variable. In mixed action space, the agent must make simultaneous discrete and continuous choices. In this section, based on the Actor-Critic framework, the PDQN-TD3 algorithm is proposed to deal with the mixed action space problem. 

\subsection{Principles of reinforcement learning}
RL is a trial-and-error-based learning method. In RL, an agent learns how to make optimal decisions by interacting with its environment. The interaction process between agent and environment can be described using Markov decision processes (MDP). The MDP model characterizes the relationship between state, action, reward, and transition probability in this process. RL formulates the optimal strategy by learning the MDP model, where the MDP can be represented as:
\begin{equation}
\label{eq:MDP}
P_{ss'}^a = P[s_{t+1}=s'|s_t=s,a_t=a]
\end{equation}
where $s_t$ represents the state at time $t$, $a_t$ represents the action taken at time $t$, $P_{ss'}^a$ represents the probability of state transition, $s$ represents the current state, $s'$ represents the next state, $a$  represents the current action, $P$ is a probability function.

In the state transition process of Eq. (\ref{eq:MDP}), a reward $R$ is generated. Given a policy $\pi$, the cumulative reward $G$ obtained by agent in the interaction process is:
\begin{equation}
\label{eq:return}
G_t =R_{t+1}+\gamma R_{t+2}+...+ \gamma^{t+k} R_{t+1+k} =  \sum_{k=0}^{\infty} \gamma^{k} R_{t+1+k}
\end{equation}
where $\gamma \in [0,1]$ is the reward discount factor, $R_{t+1}$ represents the immediate reward at time $t+1$.

The ultimate goal of the agent is to find the optimal strategy $\pi^*$ that maximizes the cumulative reward.
\begin{equation}
\label{eq:policy}
\pi^*{(s,a)} =\mathop{\rm argmax}\limits_{a} \ E[G_t]
\end{equation}
where $\pi^*(s,a)$ represents the optimal policy and $E$ represents the expectation.

To obtain the optimal strategy, Q-values are used to evaluate the superiority or inferiority of the policy  $\pi$:
\begin{equation}
\label{eq:q_pi}
Q_{\pi}(s,a) = E[R_{t}+\gamma R_{t+1}+\gamma^2 R_{t+2}+...|s_t=s,a_t=a)]
\end{equation}
Simplified further, the formula can be expressed as follows:
\begin{equation}
\label{eq:q_pi1}
Q_{\pi}(s,a) = E[R_{t}+\gamma Q(s',a')|s_t=s,a_t=a)]
\end{equation}
where $Q_{\pi}(s,a)$ represents the value of taking action $a$ in state $s$ according to policy $\pi$. 

Traditional Q-learning establishes a Q-table to store the Q-values of different actions in each state, selects the action with the maximum Q-value as the output, and updates the Q-value according to the observed reward and the next state. The Bellman equation can be expressed as:
\begin{equation}
\label{eq:Q-learning}
Q{(s,a)} = E[R + \gamma \mathop{\rm max}\limits_{a}Q(s',a')|s_t=s,a_t=a)]
\end{equation}

In practical problems, the state and action space are usually too ample to be stored in a table. Therefore, combined with deep learning, a neural network is introduced to replace the Q-table, and the output of the network is used to approximate the Q-function.
\begin{equation}
\label{eq:DQ-learning}
Q{(s,a;\theta)}  \approx Q(s,a)
\end{equation}

\subsection{PDQN algorithm}

\begin{figure}[!htb]
\centering
\includegraphics[width= 3 in]{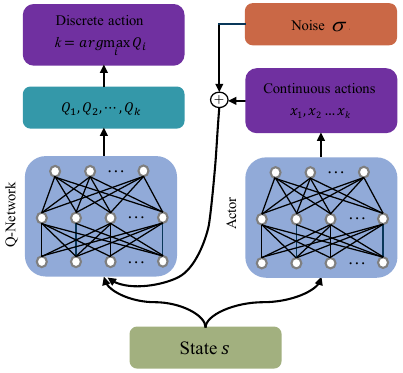}
\caption{Illustration of the PDQN architecture.}
\label{PDQN}
\end{figure}

The PDQN algorithm combines deterministic actor-critic and Q-learning, integrating the classic algorithms DDPG for continuous RL and DQN for discrete RL. Specifically, PDQN uses an actor network to output continuous actions and replaces the critic learning in deterministic actor-critic with Q-learning. This approach allows PDQN to output Q-values corresponding to each discrete action and select the discrete action based on the maximum Q-value. The schematic diagram of the PDQN algorithm is shown in Fig.\ref{PDQN}.

For PDQN, the action space $\mathcal{A}$ consists of both continuous actions and discrete actions.
\begin{equation}
\label{action}
\mathcal{A} = \left\{(k,x_k)|k \in K, x_k \in \mathcal{X}_k \right\}
\end{equation}
where $K$ denotes the discrete action set, $k$ is a discrete action, $\mathcal{X}_k$ denotes the continuous action set, $x_k$ is a continuous action.

Inspired by the processing method of DQN, the actor network of PDQN uses the deterministic policy network $x(\cdot;\mu)$ to approximate  $x_k$ to output continuous actions. The critic network approximates $Q(s,k,x_k)$ with a deep neural network  $Q(s,k,x_k;\theta)$, thus outputting discrete actions.

Eq. (\ref{eq:Q-learning}) can be further expressed as:
\begin{equation}
\label{eq:PDQN_bellman}
\begin{split}
Q(s,k,x_k) = E[R_t+{\gamma}{\ \mathop{\rm max}\limits_{k\in K}Q(s',k',x_k(s';\mu);\theta)} |s_t=s]
\end{split}
\end{equation}

The critic network parameters $\theta$ are updated based on the $TD$ error between $Q(s,k,x_k;\theta)$ and the target network estimate $y$. PDQN performs the parameter update by minimizing the loss function, which is defined as the squared error between the target Q-value and the estimated Q-value.
\begin{equation}
\label{eq:TD error}
L_Q(\theta) ={\frac{1}{2}}\Sigma(y-Q(s,k,x_k;\theta))^2 
\end{equation}
where $y = R_t+{\gamma}{\ \mathop{\rm max}\limits_{k\in K}Q(s',k',x'_k;\theta)}$.

The actor network parameters $\mu$ are updated based on the negative sum of Q values.
\begin{equation}
\label{eq:Actor network parameters}
L_x(\theta) = -\sum_{k=1}^{k}Q(s,k,x_k(s;\mu);\theta) 
\end{equation}

The target network is updated using the parameters of the actor and critic networks, and the update formula is as follows:
\begin{align}
\label{eq target network updated }
\theta_{i,targ} & \leftarrow \tau {\theta_i}+(1-\tau){\theta_{i,targ}}\\
\mu_{targ} & \leftarrow \tau {\mu}+(1-\tau){\mu_{targ}}
\end{align}

\subsection{PDQN-TD3 algorithm}

PDQN integrates DDPG and DQN, which can effectively solve continuous-discrete control problems. However, PDQN also has the corresponding drawbacks of the two algorithms. PDQN algorithm involves a maximization operation when calculating the TD target, which leads to an overestimation of the true action value by PDQN. Additionally, because the deep Q-network is continuously updated, eagerly updating the value network parameter $\theta$ when the value network is still poor fails to improve $\theta$ and destabilizes the training of the actor network due to the fluctuations in $\theta$. This paper applies a state-of-the-art PDQN-TD3 algorithm to solve the above problem. 

PDQN-TD3 uses the actor-critic network architecture. The structure of the policy network and the evaluation network are designed using the TD3 structure. The policy network includes an actor network and the corresponding target network, while the evaluation network includes two critic networks and the corresponding target network. At each time step $t$, the environment feeds the state $s$ into the actor network of PDQN-TD3 to obtain the continuous action $x_k$. The critic network acts as a Q-value network and selects a discrete action $k$ based on the state variables and the output of the actor network using an $\epsilon$-greedy policy. After executing the continuous action and discrete action $(x_k, k)$, the environment transitions to a new state $s_{t+1}$, and the tuple $(s, (x_k, k), r, s_{t+1})$ is stored in the experience replay buffer for neural network training. Compared with PDQN, the PDQN-TD3 algorithm introduces three key techniques: target policy smoothing, clipped double Q-learning, and delayed policy updates.

\subsubsection {Target policy smoothing} Random noise $\mathcal{N}(0,\sigma)$ obeying normal distribution was added to the target action value $\pi_{\mu_{targ}}(s)$ output by the actor network, and the noise value was limited within $(-c, c)$. It makes the update of the value function smooth and avoids overfitting.
\begin{equation}
\label{eq:q_noise}
\tilde{x}_k = \pi_{\mu_{targ}}(s) + clip(\mathcal{N}(0,\sigma),-c,c)
\end{equation}
where $\tilde{x}_k$ is the action value after adding smooth noise.

\subsubsection {Clipped double Q-learning} To avoid overestimating the Q-function, PDQN-TD3 introduces two independent critic networks to learn the Q-function and construct the critic computing Q-target with a smaller Q-value.
\begin{equation}
\label{eq:target Q}
Q_{targ}(s',k',\tilde{x};\theta) = {\rm min}_{i=1,2}Q_{targ}(s',k',\tilde{x};\theta_i)
\end{equation}

\begin{equation}
\label{eq:TD target}
y=r+\gamma Q_{targ}(s',k',\tilde{x};\theta)
\end{equation}

\subsubsection {Delayed policy updates} To enhance the stability of training the PDQN, the idea of delayed updates is introduced. The actor network is updated at a lower frequency, while the critic network is updated at a higher frequency. This approach ensures more stable training of the actor network.

Both critic network parameter updates are performed by minimizing the loss function, defined as the squared error between the target Q-value and the estimated Q-value.
\begin{equation}
\label{eq:TD error1}
L(\theta_1) = {\frac{1}{2}}\Sigma(y-Q(s,k,x_k;\theta_1))^2 
\end{equation}

\begin{equation}
\label{eq:TD error2}
L(\theta_2)={\frac{1}{2}}\Sigma(y-Q(s,k,x_k;\theta_2))^2 
\end{equation}

The training process of PDQN-TD3 algorithm is shown in Algorithm \ref{alg2}.

\begin{table}[!htbp]
\footnotesize
\begin{algorithm}[H]
\caption{PDQN-TD3 Algorithm Training Process.}
\label{alg2}
\begin{algorithmic}
\STATE 
\STATE  \textbf{Initialization:} Greedy algorithm coefficient $\varepsilon$ = 1, replay buffer $D$
\STATE  \textbf{Random initialization:} Online network parameters and target network parameters
\STATE  \textbf{For:} $t$ = 1: $T$ do
\STATE \hspace{0.5cm} Obtaining the initial state: $s(0)$
\STATE \hspace{0.5cm} Selecting a continuous action based on the current
\STATE \hspace{0cm} policy and exploration noise: $a_t = \pi_{\mu}(s_t) + N(0, \sigma)$
\STATE \hspace{0.5cm} Selecting a discrete action based on the exploration 
\STATE \hspace{0cm} policy: 
\begin{align}
\label{eq:opt}
\ \ k_t =
\begin{cases}
{\ \mathop{\rm argmax}\limits_{k\in K}}Q(\mu(s_t),k;\theta),   1-\varepsilon \\
{\rm rand}(K_t), \varepsilon \nonumber
\end{cases}
\end{align}
\STATE \hspace{0.5cm} Executing a($t$) in the environment, receive reward r($t$), and transmit to the next state $s(t+1)$
\STATE \hspace{0.5cm} Saving the training sample sample(t) = {s(t), a(t), r(t),
\STATE \hspace{0cm} s(t+1)} in replay buffer $D$
\STATE \hspace{0.5cm} \textbf{If:} the replay buffer $D$ data is greater than 512
\STATE \hspace{1cm} Randomly selecting a mini-batch of samples 
\STATE \hspace{0.5cm} (s(t), a(t), r(t), s(t+1)) from $D$
\STATE \hspace{1cm} Add noise to the actions in the sampled data:
\STATE \hspace{0.5cm} $\tilde{x}_k = \pi_{\mu_{targ}}(s) + clip(\mathcal{N}(0,\sigma),-c,c)$
\STATE \hspace{1cm} Calculate target value: 
\STATE \hspace{0.5cm} $y=r+ \gamma {\rm min}_{i=1,2}Q_{targ}(s',k,\tilde{x}_k;\theta_t)$
\STATE \hspace{1cm} Update the critic network:
\STATE \hspace{0.5cm} $\mu_t \leftarrow {\rm argmin}_{\mu_t}N^{-1}\Sigma(y-Q(s,k,x_k;\theta))^2$
\STATE \hspace{1cm} \textbf{If:} t mod d then:
\STATE \hspace{1.5cm} Update the actor network
\STATE \hspace{1.5cm} Update the target network
\STATE \hspace{1cm} \textbf{End if} 
\STATE \hspace{0.5cm} \textbf{End if} 
\STATE  \textbf{End for} 
\end{algorithmic}
\end{algorithm}
\end{table}

\section{Energy Management Strategy via PDQN-TD3}

\begin{figure*}[!htb]
\centering
\includegraphics[width= 6.5 in]{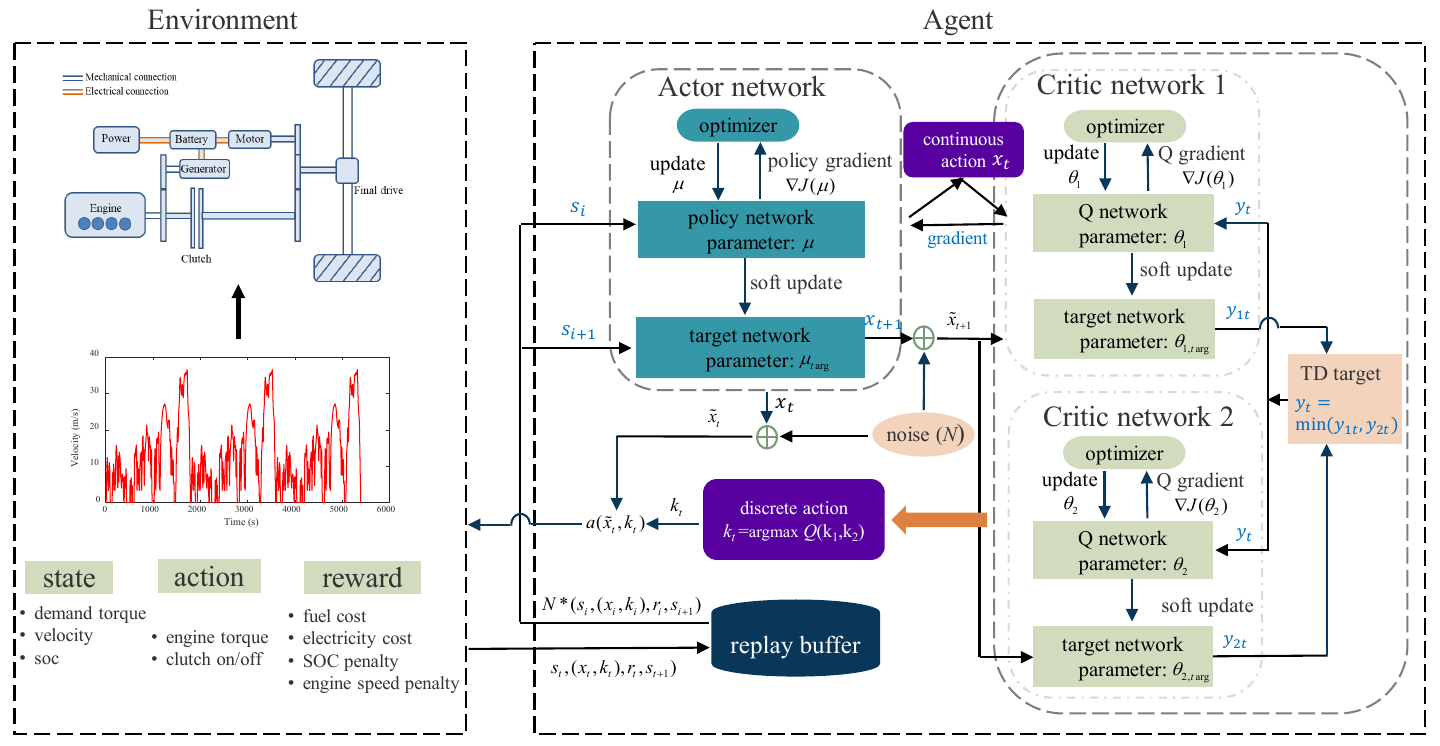}
\caption{Energy management framework based on PDQN-TD3 algorithm.}
\label{CDRL}
\end{figure*}

The framework of the PDQN-TD3 in the PHEV EMS application is depicted in Fig. \ref{CDRL}. The environment is based on the PHEV model established in section 2 and implemented in Python. The agent is the PDQN-TD3 algorithm, which takes actions based on information from the environment. At each time step, the state is fed into the actor network, which computes and outputs continuous action. Simultaneously, the continuous action and the state are used as inputs for the critic network. The critic network outputs the Q values corresponding to different discrete actions, and the discrete action is selected according to the maximum Q value.

\emph{1)} State: The state space should comprehensively reflect the environment observed by the agent. In this research, vehicle speed and demanded torque are chosen to encode the vehicle longitudinal dynamic, and SOC serves to reflect information regarding the battery's status. The state space of EMS based on PDQN-TD3 can be expressed as:
\begin{equation}
\label{eq:RL state}
\begin{split}
S = \left\{{v, T_d, SOC} \right\}
\end{split}
\end{equation}

\emph{2)} Action: The research object of this article is the DM-i series-parallel hybrid system. In this hybrid system, the PHEV series-parallel mode switching is achieved through the engagement/disengagement of the clutch, and the state of the clutch is a discrete variable. On the other hand, as indicated by Eq. (1) and (6)-(11), when the engine torque is determined, the torque of the motor and generator can be calculated based on the vehicle's demanded torque and clutch state. Therefore, the engine's output torque is selected as the continuous action and the clutch engagement/disengagement as a discrete action. The action space can be expressed as:
\begin{equation}
\label{eq:RL state}
\begin{split}
A = \left\{{T_e, k_c} \right\}
\end{split}
\end{equation}

\emph{3)} Reward: The primary goal of EMS is to minimize fuel and electricity consumption. Thus, the reward function is defined as the sum of fuel and electricity costs. To ensure that the agent does not violate the constraints of the hybrid power system during training, we also add a penalty term in the reward function for any violation of these constraints. The detailed reward function is defined as follows:
\begin{equation}
\label{total cost}
R = -(r_c+p_{\omega_e}+p_{soc})
\end{equation}
\begin{equation}
\label{r_c}
r_c = (k_f \dot{m}_f + k_e \frac{P_b}{\eta_b \eta_{chr}}) \Delta t 
\end{equation}
\begin{equation}
\label{r_{we}}
r_{\omega_e} = p_{\emph max} \Delta t 
\end{equation}
\begin{align}
\label{eq:r_{soc}}
\ \ r_{soc} =
\begin{cases}
{0}, \qquad & {\rm if}  \; SOC_l \textless SOC \textless SOC_h \\
{p_{max}\frac {SOC-SOC_h}{1-SOC_h}}, \qquad & {\rm if} \; SOC \textgreater SOC_h \\
{p_{max}\frac {SOC-SOC_l}{SOC_l}}, \qquad & {\rm if} \; SOC \textless SOC_l 
\end{cases}
\end{align}

where $p_{\omega_e}$ and $p_{soc}$ are the penalties for exceeding the constraints on the engine angular velocity and SOC, respectively. $\Delta t=1s$, $k_f$ is the fuel price which is 7.6 CNY/L, $k_e$ is the electric price which is 1.0 CNY/kW·h. $\eta_b$ is the battery efficiency, $\eta_{chr}$ is the external charger efficiency. $p_{max}=$ 0.1 is the maximum penalty, set to 10 times the maximum fuel consumption. SOC$_h$ and SOC$_l$ are the upper and lower bounds of SOC, respectively. The penalty for exceeding the SOC constraint is designed as a linear function of the SOC deviation from the desired range. Therefore, the greater the deviation of the SOC from the desired range, the higher the penalty.

\section{Simulation Experiments}

\subsection {Simulation parameters and conditions}

\begin{figure}[!htb]
\centering
\includegraphics[width= 3 in]{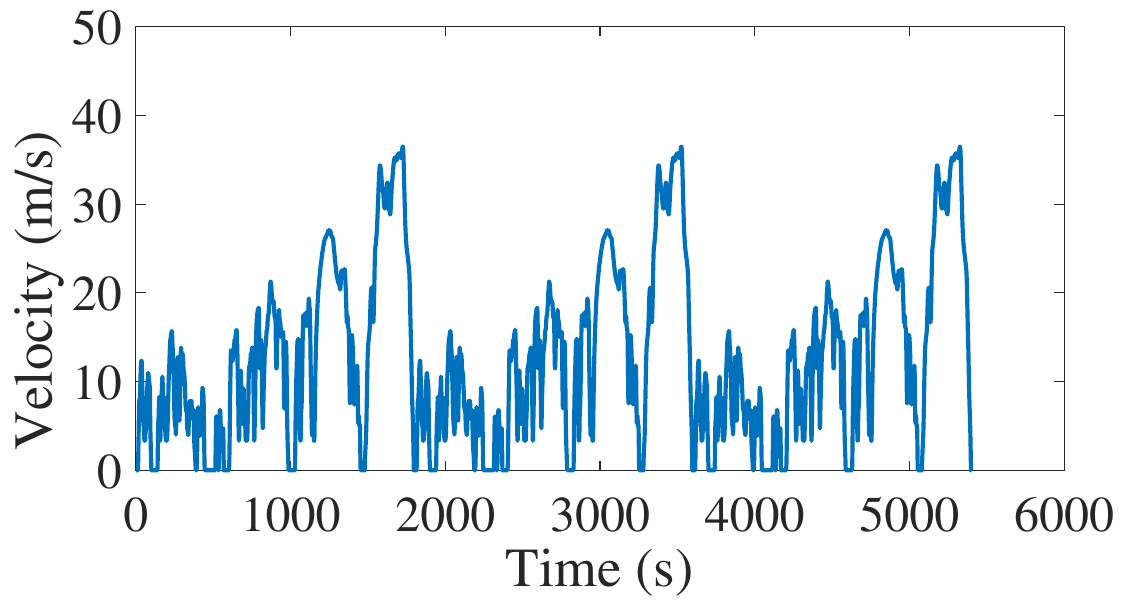}
\caption{Velocity profile of the driving cycle.}
\label{fig:wltc}
\end{figure}

In order to verify the effectiveness and superiority of the proposed EMS for PHEVs, we conducted simulation experiments using Python and compared it with EMS based on CD-CS and DP. The main hyperparameters of the PDQN-TD3 algorithm are shown in Table \ref{tab:table2}. Three WLTC (Worldwide Light-duty Test Cycle) were selected as the simulation operating conditions. WLTC is a chassis dynamometer test for determining emissions and fuel consumption from light-duty vehicles. The total time for a complete WLTC cycle is 1800s, with a driving distance of 23.25 km and a maximum vehicle speed of 120 km/h. The relationship between WLTC time and vehicle speed is shown in Fig. \ref{fig:wltc}. All the simulations were performed on a computer with an AMD Ryzen9 5950X CPU @ 3.40 GHz to ensure comparability among simulations.

The upper and lower limits of SOC are set to SOC$_h$ = 0.9 and SOC$_l$ = 0.3, respectively. During training, we introduce SOC randomization, where the initial SOC value is randomly selected in the range of 0.3 to 0.8 at the beginning of each round update. The change of SOC provides more exploration opportunities for the agent, which needs to learn the optimal energy management policy for different initial SOC values. It is important to note that the related literature shows that dynamic randomization helps agents to better deal with variable environments \cite{xie2021dynamics,andrychowicz2020learning,peng2018sim}.

\begin{table}[!htbp]
\footnotesize
\centering
\setlength{\tabcolsep}{8mm}{
\renewcommand\arraystretch{1.3} 
\caption{PDQN-TD3 parameter.}
\label{tab:table2} 
\begin{tabular}{cccc}
   \toprule [1.5 pt]
   Parameter & Value \\
   \midrule
   Soft target update  & 0.001 \\
   Reward discount factor & 0.99 \\
   Actor network learning rate & 0.0001 \\
   Q-network learning rate & 0.001\\
   Experience replay memory size & 200000 \\
   Mini-batch size & 128 \\
   Action noise & $\mathcal{N}$(0,0.02) \\
   Actor network hidden layer size & 64×64 \\
   Q-network hidden layer size & 64×64 \\
   \bottomrule
\end{tabular} }
\label{PDQN-TD3 Parameters}
\end{table}

\subsection {DP-based energy management strategy}

As a global optimization algorithm, the DP algorithm is widely employed as a benchmark for energy management. Under the premise that the entire driving cycle information is known in the future, the DP can obtain the optimal fuel economy. In order to investigate the optimization effect of the PDQN-TD3, this paper uses the DP as the benchmark for fuel economy. For DP, as the driving conditions serve as prior knowledge, the future vehicle speed and power demand are both known. At this point, SOC serves as the only state variable in the system. As the clutch switch only considers two states of engagement and disengagement, the clutch switch state is divided into two grids when performing DP. Simulation analysis shows that increasing the number of grids for engine torque and battery SOC after a certain amount does not significantly reduce the cost function of the system but does increase the computation time of the DP when the engine torque and battery SOC are discretized. Therefore, in this paper, SOC is divided into 60 grids in the range of 0.3 to 0.9, while the engine torque is divided into 120 grids in the range of 0 - 120N. In the simulation process, we use the general DP Matlab toolbox \cite{sundstrom2009generic} to obtain the benchmark.

\subsection {Rule-based energy management strategy}

The rule-based control strategy selects the optimal operation mode based on predetermined judgment conditions and control logic. It has the advantages of simplicity and easy implementation, making it a widely adopted EMS by automotive companies. To compare the fuel-saving effect of PDQN-TD3, this paper has designed a rule-based EMS. Specifically, the rule-based EMS is mainly divided into two modes: CD and CS mode. 

SOC $>$ 0.3, the vehicle enters CD mode.

\noindent (1) When the vehicle demand torque $T_d$ is greater than the engine's optimal working point:

\ding{172} If $v \geq $ 60 km/h: When $T_d$ is greater than the engine maximum working point $T_{e\_max}$, the vehicle enters the series mode. Otherwise, it enters the parallel mode.

\ding{173} If $v<$ 60 km/h, the vehicle enters the series mode.

\noindent (2) When $T_d$ is less than the engine's optimal working point:

\ding{172} If $T_d<$ 0: When SOC $>$ 0.9, mechanical braking is used; otherwise, energy is recovered by driving motor.

\ding{173} If $T_d>$ 0, the vehicle enters EV mode.

SOC $<$ 0.3, the vehicle enters CS mode.

\noindent (1) When $T_d>T_{e\_min}$:

\ding{172} If $T_d>T_{e\_max}$, then $T_e=T_{e\_max}$. If $v>$ 60 km/h, the vehicle enters the engine direct drive mode. Otherwise, it enters the series mode.

\ding{173} When $T_d<T_{e\_max}$: 
If $v>$ 60 km/h, the vehicle enters the engine direct drive mode. Otherwise, the vehicle enters the series mode.

\noindent (2) When $T_d<T_{e\_min}$: the vehicle enters the series mode.

\noindent (3) If $T_d<$ 0, energy is recovered through the driving motor.

For the above rules:

(1) In series mode, the clutch is disengaged, and the engine operates within its optimal economic range to drive the generator for power generation while the motor provides the demand torque.

(2) In parallel mode, the clutch is engaged, the generator does not work, and the motor aids the engine to drive the vehicle.

(3) In EV mode, the clutch is disengaged, neither the engine nor the generator works, and the motor provides the demand torque.

(4) In engine direct drive mode, the clutch is engaged, the generator does not work, and the engine directly drives the vehicle.

\subsection {Simulation results analysis}

\begin{figure}[!htb]
\centering
\includegraphics[width= 3 in]{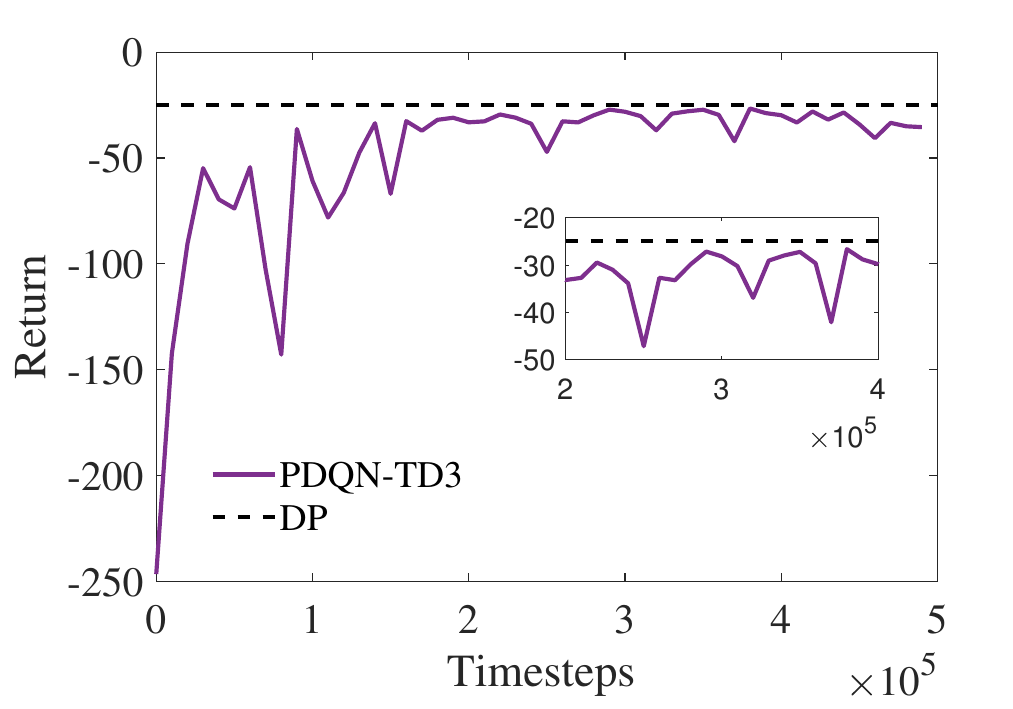}
\caption{Episode reward sum in training.}
\label{return}
\end{figure}

As shown in Fig. \ref{return}, the curve represents the cumulative reward changes, where a higher return value indicates a better learning effect.  It can be observed that the return value curve fluctuates but exhibits an overall upward trend, indicating that the intelligent agent continuously adjusts its strategy to maximize the cumulative return per episode.  After $17 \times 10^4$ steps of iteration, the algorithm gradually converges to the optimal control strategy.  The PDQN-TD3 and the DP-based strategy exhibit similar total energy consumption, suggesting that the PDQN-TD3 control strategy has better fuel economy.

\begin{figure}[!htb]
\centering
\includegraphics[width= 3 in]{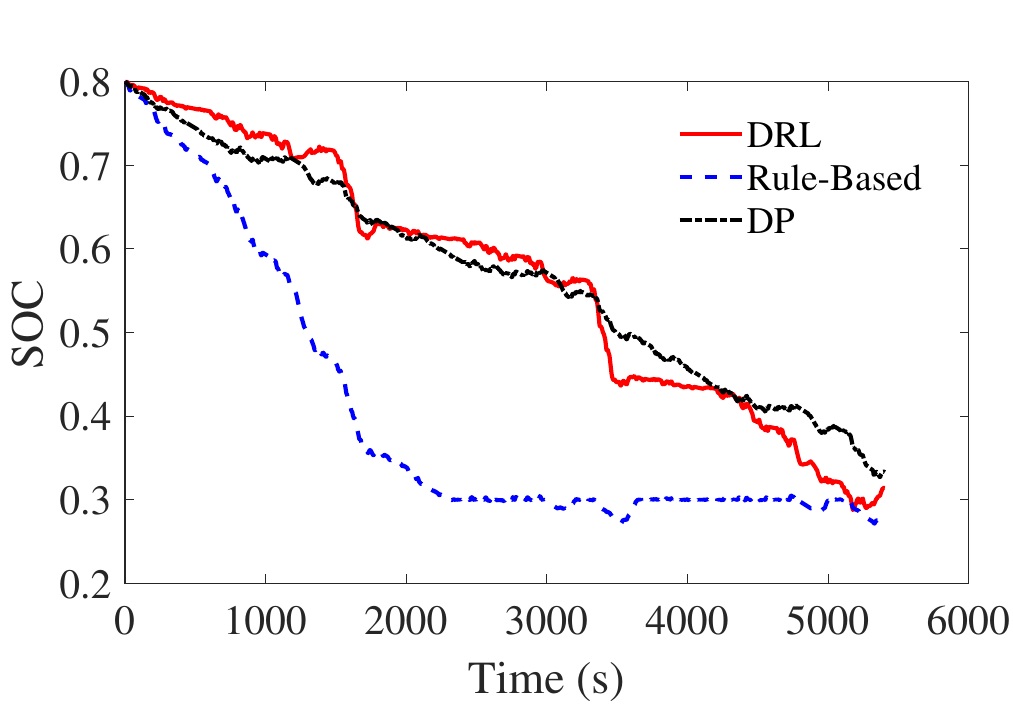}
\caption{Battery SOC trajectories under different EMSs.}
\label{SOC}
\end{figure}

The comparison of the vehicle SOC variation trends over time for three algorithms under the WLTC cycle is shown in Fig. \ref{SOC}. It can be observed that the CD-CS strategy tends to utilize electric energy rather than fuel when the SOC is greater than $\rm SOC_l$. When the SOC drops to the set threshold, it fluctuates around $\rm SOC_l$, and the engine becomes the main power source to suppress excessive battery discharge. However, using this control strategy under aggressive driving conditions requires the engine to provide significant torque output when the battery is depleted, reducing the vehicle's fuel economy. On the other hand, the variation trend generated by the PDQN-TD3 and DP algorithms is very similar, but the SOC decrease is more gradual with PDQN-TD3. This indicates that when PDQN-TD3 performs energy management, the engine startup frequency to charge the battery increases compared to DP under the same driving conditions. This approach reduces the peak discharge power of the power battery, which is positive and beneficial for the battery's lifespan. Compared with CD-CS, the motor can provide greater output power under aggressive driving conditions, and the engine working point can be adjusted more reasonably, allowing the engine to work more often within the optimal fuel economy range.

Table \ref{comparison CDCS cost} shows the simulation results of the total system energy consumption for fuel and electricity consumption. As shown in Table \ref{comparison CDCS cost}, the DP-based achieves optimal fuel economy, which we use as a benchmark for comparison with the PDQN-TD3 and CD-CS methods. Under the charge-depleting state, the PDQN-TD3 EMS improves energy efficiency by 8.3\% compared to the CD-CS EMS. The fuel economy gap between PDQN-TD3 EMS and DP is 6.6\%. In the charge-sustaining state, the PDQN-TD3 EMS strategy outperforms the CD-CS strategy by 4.1\% in terms of fuel economy, with a difference of approximately 3.9\% compared to DP. These results indicate that PDQN-TD3 has better energy utilization efficiency, which reduces the operating costs of HEVs, further demonstrating the optimality of the PDQN-TD3 strategy for energy management. 

\begin{table}[!t]
\footnotesize
  \centering
  \caption{Comparison analysis of total cost in different methods.}
  \begin{tabular}{lllcl}
    \toprule
    \multirow{2}{*}{SOC$_{initial}$} & Algorithm  &  Cost & Fuel consumption & Gap \\
                        &            &  (CNY)     & (L)         & \\
    
    \midrule
    & PDQN-TD3  & 26.61  & 3.05  & 6.6\% \\
    0.8 & CD-CS     & 29.03    & 3.28  & 16.3\% \\
    & DP        & 24.96   & 2.86  & 0 \% \\

    \midrule
    & PDQN-TD3  & 31.50    & 4.14  & 3.9\% \\
    0.3 & CD-CS     & 32.85     & 4.34  & 8.3\% \\
    & DP        & 30.32     & 4.05  & 0 \% \\
    \bottomrule
  \end{tabular}
  \label{comparison CDCS cost}
\end{table}

\begin{figure*}[!t]
  \centering
  \subfigure[]{\includegraphics[width=0.32\textwidth]{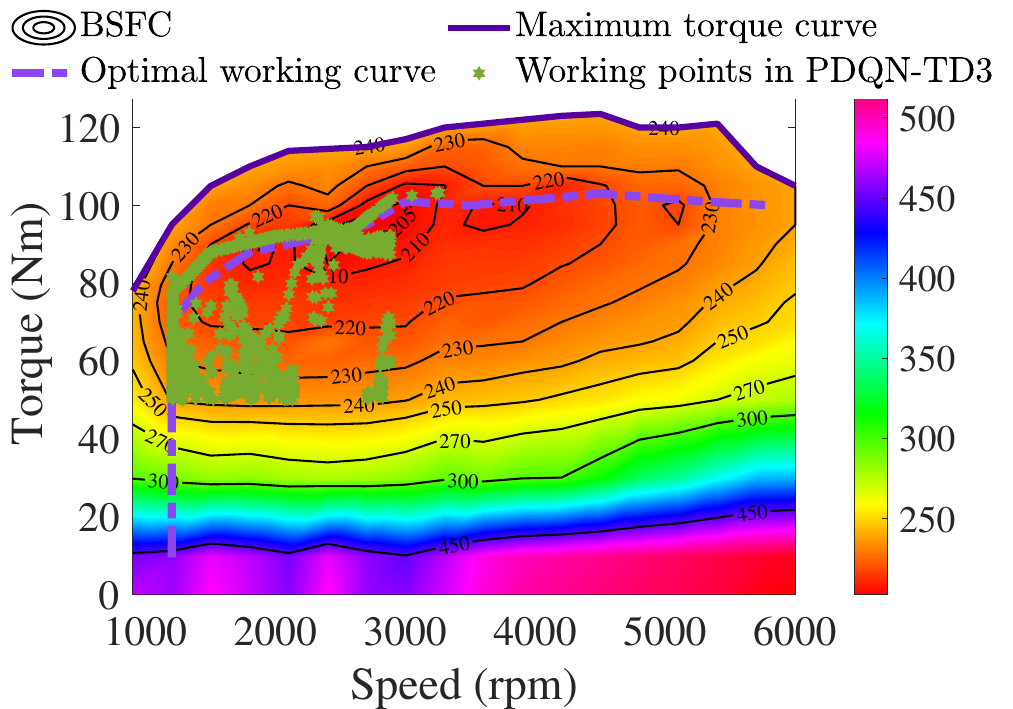}}
  \subfigure[]{\includegraphics[width=0.32\textwidth]{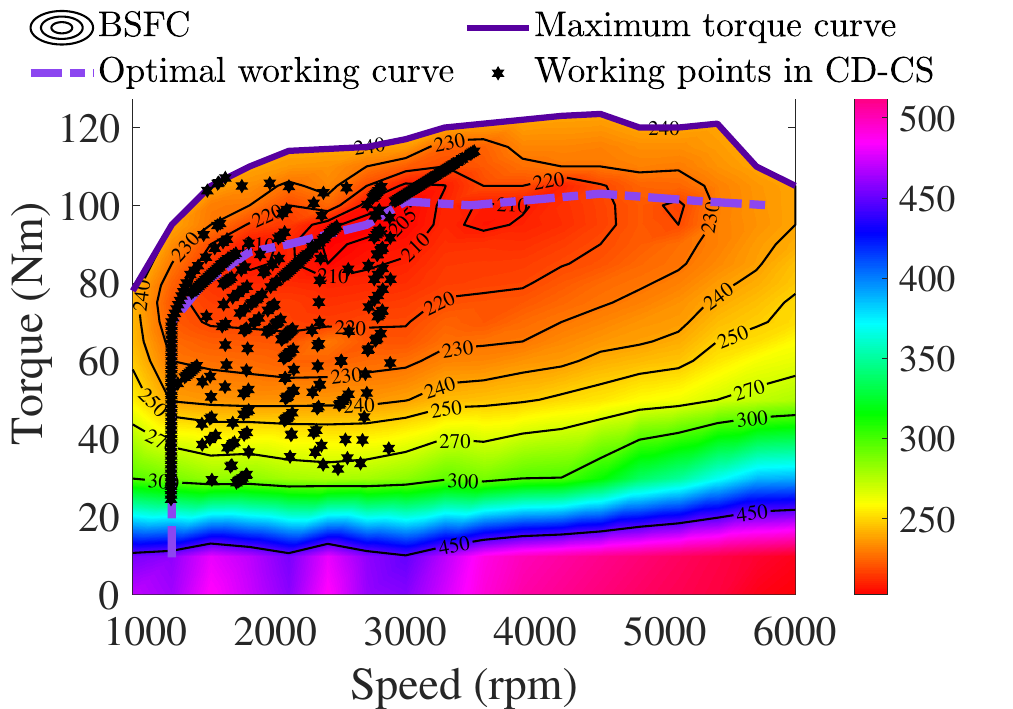}} 
  \subfigure[]{\includegraphics[width=0.32\textwidth]{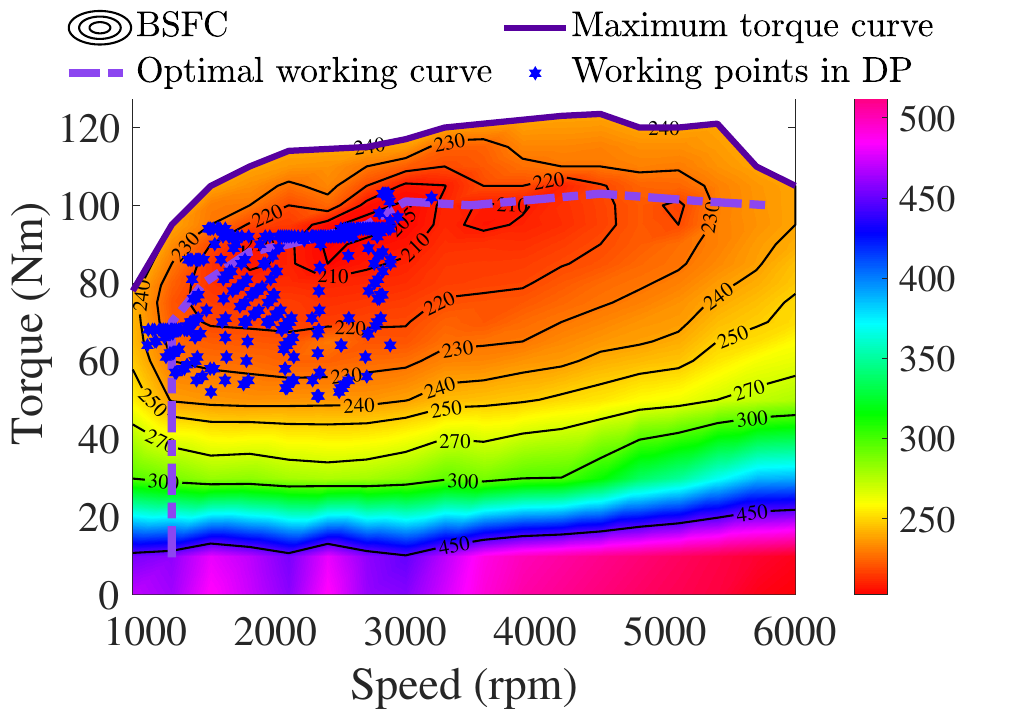}} 
  \caption{(a) Engine working points of PDQN-TD3. (b) Engine working points of CD-CS. (c) Engine working points of DP.
  }
  \label{test_dp_rl_cdcs}
\end{figure*}

Fig. \ref{test_dp_rl_cdcs} shows the engine working points for the three control strategies. It can be observed that the EMS based on DP has more engine working points within the fuel-efficient range. The main reason for the sparse distribution of engine working points is that DP discretizes the engine torque. Due to the influence of the discretization precision, the engine working point can only be selected from discrete intervals and limited discrete engine operating points on the map. Therefore, one point on the map may correspond to many engine torques with the same value. While DP has better optimization results, it has a longer computation time than PDQN-TD3 and CD-CS, which is the main reason why the DP algorithm cannot be applied to real vehicles. Compared with the EMS based on CD-CS, the control strategy based on PDQN-TD3 has smaller fluctuations in engine speed and torque, which further indicates that the EMS based on PDQN-TD3 can adjust the engine operating point well so that the engine can work in the optimal fuel economy zone in most cases. Compared with CD-CS, the engine operates more efficiently and has better fuel economy. The engine in CD-CS works in a non-economical range because when the SOC is lower than $\rm SOC_l$, the output power of the battery is restricted and may even be unable to provide power to the vehicle. During such times, the vehicle must rely solely on the engine, which can limit engine efficiency and reduce fuel economy.

\begin{figure*}[!t]
  \centering
  \subfigure[]{\includegraphics[width=0.32\textwidth]{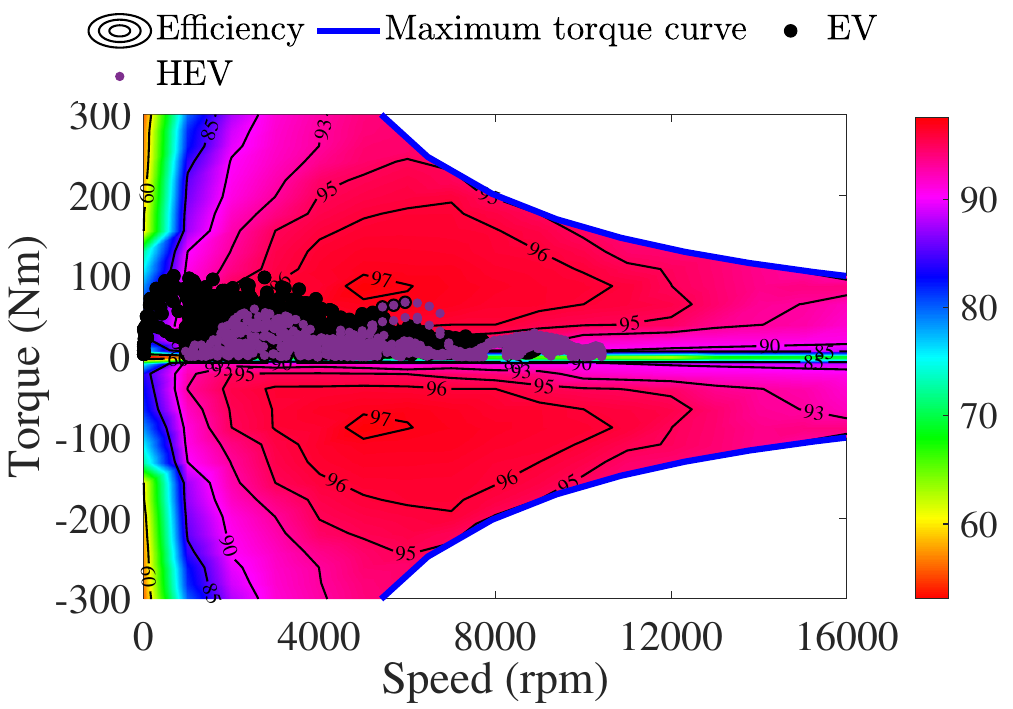}}
  \subfigure[]{\includegraphics[width=0.32\textwidth]{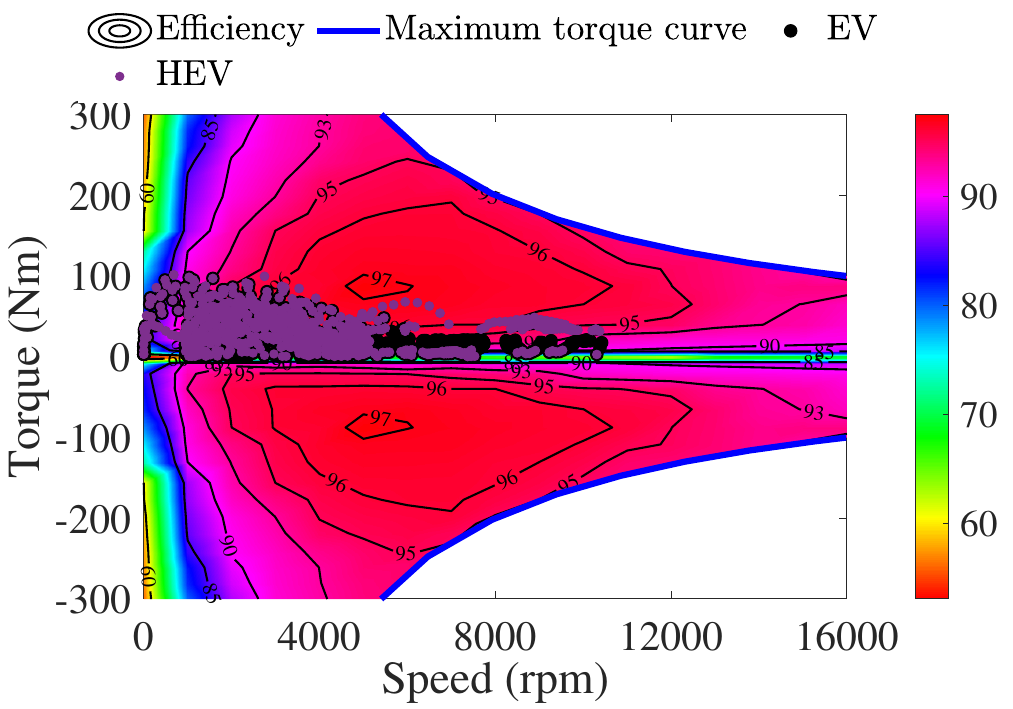}} 
  \subfigure[]{\includegraphics[width=0.32\textwidth]{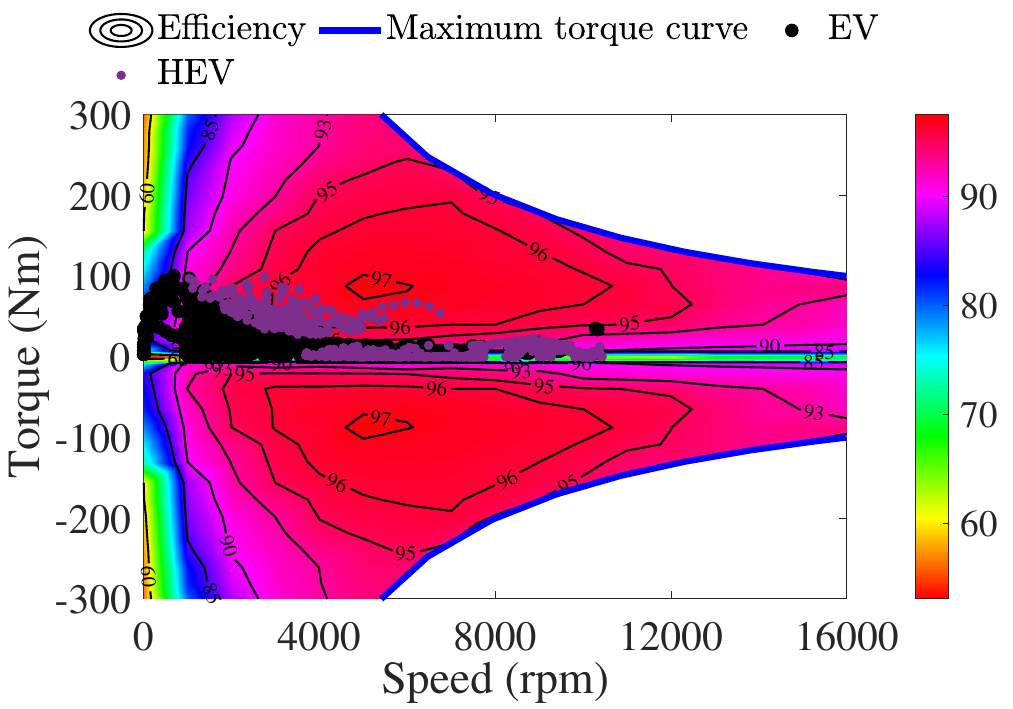}} 
  \caption{(a) Motor working points of PDQN-TD3. (b) Motor working points of CD-CS. (c) Motor working points of DP.
  }
  \label{test_motor_dp_rl_cdcs}
\end{figure*}

Fig. \ref{test_motor_dp_rl_cdcs} shows the working points of the three control strategies for the electric motor. It can be seen that for PDQN-TD3 and DP algorithms, the EV mode is mainly distributed in the low area vehicle speed range. As the vehicle speed increases, the engine starts, and the system operates in HEV mode. Unlike DP and PDQN-TD3,  the CD-CS control strategy frequently enters HEV mode even at very low vehicle speeds. This is because when the battery power is depleted, the battery cannot meet the power demand of the system and relies on the engine generation to provide additional power frequently. However, DP and PDQN-TD3 can plan battery usage more reasonably by driving the vehicle using the engine at higher speeds. This not only improves fuel economy but also reduces the motor's output torque, which is beneficial for the battery's lifespan.

It is worth noting that for SOC = 0.3, we reach similar conclusions as for SOC = 0.8. Therefore, the simulation results for the engine and motor operating points with SOC=0.3 are no longer shown in this paper.

The statistical results of the clutch engagement or disengagement are presented in Table \ref{tab:3}. It can be seen that, for all three control strategies, the clutch disengagement times are greater than the clutch engagement times. This is attributed to the lower cost of electricity compared to fuel. Although direct drive of the engine is more economical when the clutch is engaged, it cannot guarantee that the engine operates in the economic range at low speed or low torque. Instead, it increases fuel consumption. In this case, the system tends to generate electricity through the engine in the high-efficiency range instead of directly driving the vehicle. The difference between the three strategies is that the CD-CS has significantly fewer clutch engagements. This is because at the beginning of the journey, CD-CS tends to use electricity, and the engine is almost not started until the system's torque demand or vehicle speed exceeds the set threshold. When the battery SOC is low, although the engine needs to be started frequently to provide torque, the system does not have a transmission, and the engine cannot directly drive the vehicle at any speed. More often, the engine acts as a range extender. The clutch engagement percentages in PDQN-TD3 and DP are close but still have a gap, possibly due to the lower fuel efficiency of PDQN-TD3 than DP. PDQN-TD3 cannot predict the operating conditions of the entire journey in advance like DP, which may limit the optimization effect of PDQN-TD3 to some extent.

\begin{table}[!t]
\footnotesize
  \centering
  \caption{The statistical results of the clutch engagement/disengagement.}
  \begin{tabular}{crc}
    \toprule
    SOC$_{initial}$       & Algorithm  &  Clutch engagement percentage             \\
    \midrule
    \multirow{3}{*}{}\\  & PDQN-TD3         & 18.0\%                         \\
  0.8                 & CD-CS           & 8.5\%                    \\
                             & DP             &19.3\%                    \\  %
    \midrule
    \multirow{3}{*}{}\\  & PDQN-TD3          & 25.2\%                                    \\
  0.3                  & CD-CS               & 15.7\%                        \\  %
                             & DP                 & 21.4\%                         \\  
    \bottomrule
  \end{tabular}
  \label{tab:3}
\end{table}

\subsection {Generalization performance}
To evaluate the generalization performance of the PDQN-TD3 algorithm in PHEV EMS, we compare the fuel economy performance of the PDQN-TD3 algorithm under two different driving conditions: the New European Driving Cycle (NEDC) and the Comprehensive Long-Term Cycle (CLTC). The results of the generalization tests have been presented in Table \ref{tab:generalization}.

\begin{table}[!t]
\footnotesize
  \centering
  \caption{Generalization results with the new cycles.}
  \begin{tabular}{llcr}
    \toprule
    Test cycle               & Strategy  &  total cost/CNY & gap             \\
    \midrule
    \multirow{3}{*}{5*NEDC}\\  & PDQN-TD3  & 22.19           & 9.85\%                         \\
  (SOC$_{initial}$=0.8)      & CD-CS     & 23.39          & 15.79\%                    \\
                             & DP        & 20.20           &0\%                    \\  
    \midrule
    \multirow{3}{*}{5*NEDC}\\  & PDQN-TD3  & 26.37           & 4.68\%                                    \\
  (SOC$_{initial}$=0.3)      & CD-CS     & 27.64           & 9.73\%                        \\  %
                             & DP        & 25.19           & 0\%                         \\  
    \midrule
    \multirow{3}{*}{3*CLTC}\\  & PDQN-TD3  & 17.04           & 10.08\%                                     \\
   (SOC$_{initial}$=0.8)     & CD-CS     & 18.01           & 16.34\%                                    \\
                             & DP        & 15.48           & 0\%                        \\  
    \midrule
    \multirow{3}{*}{3*CLTC}\\  & PDQN-TD3  & 21.35           & 4.30\%                               \\
      (SOC$_{initial}$=0.3)  & CD-CS     & 22.57           & 10.26\%                                 \\
                             & DP        & 20.47           & 0\%                       \\  
    \bottomrule
  \end{tabular}
  \label{tab:generalization}
\end{table}

Our generalization performance evaluation focuses on the fuel consumption of the EMS under different operating conditions. As shown in Table \ref{tab:generalization}, it can be observed that, even in new driving cycles, regardless of whether the initial SOC is high or low, PDQN-TD3 exhibits better fuel economy than CD-CS. In the case of PHEV low SOC, the minimum gap between the proposed strategy and DP is merely 4.3\%, the maximum gap is just 10.08\%, while CD-CS has a maximum difference of 16.34\%.

\section{Conclusions}

This work focuses on the BYD DM-i hybrid system and conducts mathematical modeling of the hybrid system and EMS optimal control. The main conclusions are as follows:

Considering the characteristics of the hybrid system with both continuous and discrete variables, we establish a control-oriented model for the DM-i hybrid systems from the perspective of mixed-integer programming. This approach enables the simultaneous handling of both continuous and discrete variables in energy management.

The CDRL algorithm PDON-TD3 was applied to EMS, achieving simultaneous optimization of both continuous and discrete actions. We introduce SOC randomization during training to ensure the algorithm's generalization performance. The PDQN-TD3 EMS has better fuel economy effects than CD-CS, with a 8.3\% and 4.1\% reduction in the total cost of fuel consumption and electric energy consumption for high and low SOC, respectively. The cost-effectiveness gap is 6.6\% and 3.9\% compared with DP  for high and low SOC, respectively.

Future research can explore integrating prediction mechanisms into RL algorithms, further improving the energy utilization efficiency and driving comfort of HEVs. In addition, the cooperative optimization of EMS and advanced driving assistance systems for HEV is also possible future work for energy saving and emission reduction. 

\section*{Acknowledgement}\label{}
This work is supported by Guangzhou Basic and Applied Basic Research Program under Grant 2023A04J1688.
\printcredits

\section*{Conflict of Interest}\label{}
The authors declare no conflict of interest. 
\printcredits

\section*{Data Availability Statement}\label{}
The data that support the findings of this study are available from the corresponding authors upon reasonable request.
\printcredits

\bibliographystyle{unsrt}
\bibliography{Reference}

\end{sloppypar}
\end{document}